\documentclass[%reprint,
superscriptaddress,
twocolumn,
 amsmath,amssymb,
 aps,
 prl,
]{revtex4-2}

\usepackage{color}

\usepackage{graphicx}% Include figure files
\usepackage{dcolumn}% Align table columns on decimal point
\usepackage{bm}% bold math
\usepackage{hyperref}% add hypertext capabilities
\usepackage{fancyhdr}
\newcommand{\diff}{\textrm{d}}

\begin{document}

\newcommand{\RHheaderline}{\textsf{P3H-26-028, TTK-26-07, IFJPAN-IV-2026-11, Cavendish-HEP-26/01}}
%%----------------------------------------------------------------------
\fancypagestyle{firstpage}
{
  \renewcommand{\headrulewidth}{0pt}
  \fancyhead[R]{\RHheaderline}
}

\title{Next-to-next-to-next-to-leading order QCD corrections to photon-pair production}% Force line breaks with \\

\author{Micha\l{ }Czakon}%
 \email{mczakon@physik.rwth-aachen.de}%
 \affiliation{Institute for Theoretical Particle Physics and Cosmology, RWTH Aachen University, 52056 Aachen, Germany.}%
\author{Felix Eschment}%
 \email{felix.eschment@rwth-aachen.de}%
 \affiliation{Institute for Theoretical Particle Physics and Cosmology, RWTH Aachen University, 52056 Aachen, Germany.}%
\author{Terry Generet}%
 \email{generet@hep.phy.cam.ac.uk}%
 \affiliation{Cavendish Laboratory, University of Cambridge, Cambridge CB3 0HE, United Kingdom}%
\author{Rene Poncelet}
 \email{rene.poncelet@ifj.edu.pl}
\affiliation{Institute of Nuclear Physics PAN, ul.\ Radzikowskiego 152, 31-342 Krak\'ow}

\date{\today}% It is always \today, today,
             %  but any date may be explicitly specified

\begin{abstract}
The production of two isolated photons at the Large Hadron Collider has applications in Higgs-boson phenomenology and in new-physics searches. In fact, measurements would provide more insight if a percent-level theoretical prediction were available. We account for next-to-next-to-next-to-leading order strong-interaction effects in the cross section for the first time. This is not only necessary to demonstrate the convergence of perturbation theory, but also results in a reduction of the theoretical uncertainty by a factor of three down to 3\%. With this, we pave the way to improved analyses.
\end{abstract}

%\keywords{Suggested keywords}%Use showkeys class option if keyword
                              %display desired
\maketitle
\thispagestyle{firstpage}

%\tableofcontents

\section{Introduction}

Photons are among the cleanest objects at hadron colliders: they deposit their energy in the electromagnetic calorimeters, can be selected efficiently by dedicated photon triggers, and yield diphoton spectra with excellent invariant-mass, $m_{\gamma\gamma}$, resolution. In fact, the Higgs-boson decay channel into two photons was instrumental in establishing discovery. The same channel remains a source of highly accurate data on the Higgs-boson transverse-momentum and rapidity distributions \cite{ATLAS:2022fnp,CMS:2025fwn}, and, not the least, associated jet activity. Most importantly, it allows for measurements of the mass \cite{ATLAS:2023owm,CMS-PAS-HIG-24-007} of the fundamental scalar with competitive uncertainties. The same clean invariant-mass spectrum that made $H \to \gamma\gamma$ powerful also makes diphoton production a prime channel for new-particle searches. For instance, it provides strong constraints on a multitude of fashionable models at high $m_{\gamma\gamma}$ \cite{ATLAS:2021uiz,CMS:2024nht}, whereas low-$m_{\gamma\gamma}$ measurements are being conducted in order to exclude or discover dark-matter candidates with masses below the electroweak scale \cite{ATLAS:2022abz,CMS:2026zsp}. The high-luminosity phase of the LHC (HL-LHC) promises more measurements with higher statistics and improved systematics due to the constant methodological progress on the experimental-analysis side.

How important is reliable theoretical modeling of prompt diphoton production---the background process to the signals listed above? The complexity of Monte Carlo event generators and the numerous approximations they rely on, make it desirable to assess the quality of modeling in the context of purely perturbative predictions. Consider the $H \to \gamma\gamma$ signal as an example. The signal excess over the continuum background is about 4\% in the main $m_{\gamma\gamma}$ window, see e.g., Ref.~\cite{ATLAS:2023owm}. In order to perform the analysis, the background is actually fitted as a smooth curve from the region outside the signal window. Ref.~\cite{ATLAS:2022fnp} quotes an uncertainty of about 4\% on the inclusive $pp \to H \to \gamma\gamma$ cross section from uncertainty on the shape of the fit function alone as one of the largest systematic errors. This is where an accurate theory prediction could help, especially if percent-level accuracy would be certified. Cross-section measurements are, in fact, performed in a multitude of specialized phase-space regions adapted to the goals of a given analysis. This fine-graining will be even more pronounced at the HL-LHC. A simple fit-function is rarely applicable in such cases and the shape of the background is obtained from Monte Carlo simulations with the normalization given from inclusive fits. Clearly, a percent-level perturbative prediction would be invaluable in validating the simulations. One should also remember that the concept of background diphoton production ignores quantum interference effects between signal and background amplitudes. These percent-level effects are not negligible \cite{Bargiela:2022dla}. However, their inclusion is only justified if the background process without interference is known to matching precision.

Perturbative predictions for diphoton production present a major challenge. Inclusion of strong-interaction effects at next-to-leading (NLO) and next-to-next-to-leading (NNLO) \cite{Catani:2011qz, Campbell:2016yrh} showed that while theory starts to agree with the data at NNLO \cite{ATLAS:2021mbt}, no convergence is observed when increasing the order of the approximation. Instead, both NLO and NNLO calculations yield cross-section values well outside the theory uncertainty predicted by the previous order. Furthermore, this theory uncertainty increases order-by-order and reaches about 8\% at NNLO. The reason for the unsatisfactory behavior lies to a large extent in the fact that new parton-scattering channels appear at NLO (quark-gluon) and NNLO (gluon-gluon), while the leading order is generated by electromagnetic quark-anti-quark annihilation. Hence, reliability of the perturbative expansion can only be assessed with results beyond NNLO.

In this letter, we present the theoretical prediction for the fiducial cross section for diphoton production at next-to-NNLO ($\text{N}^3$LO) in QCD. We finally observe perturbative convergence together with a significant reduction of scale uncertainty to about 2\%. There are two other sources of uncertainty that affect our result. One is due to Monte Carlo integration, while the other is an artifact of the computational method, and is referred to as power corrections. We estimate both uncertainties at the same level of 2\%.

Our result is first in the class of genuine two-to-two scattering processes at hadron colliders. It could not have been obtained without methodological progress, despite substantial computational resources used. In particular, we have drastically improved the efficiency of the NNLO $pp \to \gamma\gamma$ + jet calculation, and obtained analytic one-loop amplitudes for $pp \to \gamma\gamma$ + 2 jets using techniques that have only recently been proposed.

\section{Calculation}

$\text{N}^3$LO predictions for simpler hadron-collider processes have been obtained with a multitude of methods: reverse unitarity \cite{Anastasiou:2002yz, Anastasiou:2002wq} in Refs.~\cite{ Anastasiou:2015vya, Anastasiou:2016cez, Mistlberger:2018etf, Duhr:2019kwi, Duhr:2020seh, Duhr:2020sdp, Baglio:2022wzu}, projection-to-Born \cite{Cacciari:2015jma} in Refs.~\cite{ Chen:2021isd, Dreyer:2016oyx, Dreyer:2018qbw}, $q_T$-slicing \cite{Catani:2007vq} in Refs~\cite{Cieri:2018oms, Billis:2021ecs, Camarda:2021ict, Chen:2021vtu, Camarda:2021jsw, Chen:2022cgv, Chen:2022lwc, Neumann:2022lft, Campbell:2023lcy}. Since there is no subtraction scheme able to handle our problem~\footnote{See, however, Ref.~\cite{Chen:2025kez} for recent progress in antenna subtraction \cite{Gehrmann-DeRidder:2005btv} at N${}^3$LO for $e^+e^-$ collisions.}, we apply the $q_T$-slicing method~\cite{Catani:2007vq}, which, at this order, is currently restricted to colorless final states. Approaches~\cite{Buonocore:2022mle, Buonocore:2023rdw, Fu:2024fgj, Buonocore:2025ucn, Gao:2026tnd, Dong:2026eas} to overcome this limitation have not exceeded NNLO so far~\footnote{Notice, however, the recent progress for lepton-hadron collisions in Ref.~\cite{Dong:2026uvw}.}. A different slicing approach that could also be applied is $N$-jettiness slicing~\cite{Boughezal:2015dva, Gaunt:2015pea, Ebert:2020unb, Baranowski:2022vcn, Bruser:2018rad, Banerjee:2018ozf}. However, it was only completed \cite{Baranowski:2024vxg, Baranowski:2024ysi} for our case, when the present study was already under way.

In the framework of $q_T$-slicing, the diphoton production cross section is decomposed as
\begin{equation}
\label{eq:qt-slicing}
    \sigma^{pp\to \gamma\gamma}_{\text{N}^3\text{LO}} = \int_0^{q_{T}^{\text{cut}}} \diff q_T \frac{\diff \sigma^{pp\to \gamma\gamma}_{\text{N}^3\text{LO}}}{\diff q_T} + \int_{q_{T}^{\text{cut}}}^{\infty} \diff q_T \frac{\diff \sigma^{pp\to \gamma\gamma j}_{\text{NNLO}}}{\diff q_T},
\end{equation}
where $q_T\equiv|\bm{q}_T|$ is the transverse momentum of the photon pair. A finite $q_T^\text{cut}>0$ regulates infrared divergences due to the jet radiation in the second term. On the other hand, if $q_T^\text{cut}$ is chosen sufficiently small, we can approximate the first term by means of the factorization theorem~\cite{Collins:1984kg,Bozzi:2005wk,Becher:2010tm,Echevarria:2011epo,Chiu:2012ir}
\begin{multline}
    \frac{\diff \sigma^{pp\to\gamma\gamma}}{\diff \bm{q}_T^2 \, \diff m_{\gamma\gamma} \, \diff y} = \sum_{a,b} H_{ab}(m_{\gamma\gamma}) \int \!\frac{\diff^2 \bm{b}_T}{(2\pi)^2} e^{i\bm{q}_T\cdot \bm{b}_T} \\
    \times \tilde{f}_a(x_1,\bm{b}_T) \tilde{f}_b(x_2,\bm{b}_T) + \mathcal{O}(q_T/m_{\gamma\gamma}),
\end{multline}
where $m_{\gamma\gamma}$ is the diphoton invariant mass and $y$ the rapidity, $x_{1,2} = m_{\gamma\gamma}/\sqrt{s} \exp(\pm y)$. $H_{ab}$ contains the Born cross section and virtual corrections and $\tilde{f}_i$ is the transverse-momentum-dependent parton distribution function. It combines beam and soft functions, $\tilde{f}_i(x,\bm{b}_T) \equiv B_i(x,\bm{b}_T)\sqrt{S(\bm{b}_T)}$, and enters multiplicatively for the partons $a$ and $b$ when expressed through the impact parameter $\bm{b}_T$, the Fourier conjugate of $\bm{q}_T$. Furthermore, it is related to regular parton distribution functions (PDFs) $f_i$ via convolution~\cite{Collins:1984kg}
\begin{equation}
    \tilde{f}_i(x,\bm{b}_T) = \sum_j \int_x^1\! \frac{\diff z}{z} \tilde{\mathcal{I}}_{ij}^\text{TMD}(z,\bm{b}_T) f_j\Big(\frac{x}{z}\Big) +\mathcal{O}(\bm{b}_T\Lambda_\text{QCD}),
\end{equation}
with the matching kernel $\tilde{\mathcal{I}}_{ij}^\text{TMD}$ calculated through $\text{N}^3$LO in Refs.~\cite{Ebert:2020yqt,Luo:2020epw}. Starting from the ancillary files of Ref.~\cite{Luo:2020epw}, we perform the $\bm{b}_T$ and $\bm{q}_T$ integrals analytically and take the virtual amplitudes through three-loop order that contribute to $H_{ab}$ from Refs.~\cite{Glover:2003cm, Caola:2020dfu}. 

The second term of Eq.~\eqref{eq:qt-slicing} requires an NNLO calculation that can, in principle, be performed with existing subtraction schemes~\cite{Gehrmann-DeRidder:2005btv, Daleo:2006xa, Currie:2013vh, NNLOJET:2025rno, Czakon:2010td, Somogyi:2006cz, DelDuca:2016ily, DelDuca:2024ovc, Caola:2017dug, Caola:2019nzf, Klein:2026tlj, Magnea:2018hab}. In particular, Refs.~\cite{Gehrmann-DeRidder:2005btv, Czakon:2010td} have enabled precise predictions for $2\to 3$ processes \cite{Chawdhry:2019bji, Czakon:2021mjy, Chen:2022ktf, Hartanto:2022qhh, Alvarez:2023fhi, Badger:2023mgf, Badger:2025ilt} including $pp \to \gamma\gamma j$ \cite{Chawdhry:2021hkp, Buccioni:2025bkl}. Here, we use \texttt{STRIPPER}, a \texttt{C++} implementation of the NNLO sector-improved residue subtraction scheme~\cite{Czakon:2010td, Czakon:2014oma, Czakon:2019tmo} starting from the setup described in detail in Ref.~\cite{Chawdhry:2021hkp}. Unfortunately, the requirement of a very small $q_T^\text{cut}$ parameter introduces major technical obstacles. We therefore implement several improvements compared to Ref.~\cite{Chawdhry:2021hkp}.
First, we employ hybrid floating-point precision, double or quadruple, using the \texttt{qd} library~\cite{qd-library}. Phase-space regions with particularly large cancellations are calculated in quadruple precision by default.
Second, a significant speedup of the Monte Carlo integration was achieved by modifying, as described in the supplementary materials 
\cite{suppl}, the selector functions, which are integral parts of the \texttt{STRIPPER} methodology.
Third, we use amplitudes of very high numerical stability. The seven-point tree-level amplitudes are computed with the \texttt{avhlib} library~\cite{avhlib, Bury:2015dla} and the five-point two-loop amplitude from Ref.~\cite{Chawdhry:2021hkp} is upgraded to full-color thanks to Ref.~\cite{Agarwal:2021vdh}. The previously used \texttt{OpenLoops} library~\cite{Cascioli:2011va, Buccioni:2019sur} does not offer sufficient precision for the six-point one-loop amplitudes. While the library \texttt{NJet}~\cite{Badger:2012pg} can calculate one-loop amplitudes as needed with octuple precision, the computation time makes a calculation with acceptable statistical precision unfeasible. Therefore, we implement stable and compact analytic expressions for the amplitudes, see next section.

Note that the cross section contains an independently gauge-invariant contribution from the loop-induced $gg\to \gamma \gamma$ amplitude at NNLO. The $gg\to \gamma \gamma g$ amplitude relevant at $\text{N}^3$LO is obtained through \texttt{OpenLoops}. Since this loop-induced $\text{N}^3$LO contribution is obtained with an effective NLO calculation, which had already been performed in Ref.~\cite{Campbell:2016yrh}, the precision is sufficient in this case. With the three-loop $gg\to\gamma\gamma$~\cite{Bargiela:2021wuy} and two-loop $gg\to \gamma \gamma g$~\cite{Badger:2021imn} amplitudes, all ingredients are available to include this contribution at $\text{N}^4$LO. However, we choose not to do so in order to obtain a pure $\text{N}^3$LO prediction.

We work in a theory with $n_f=5$ flavors without top-quark loops and a QED coupling of $\alpha=1/137$. The strong coupling and PDF values are obtained through the LHAPDF interface~\cite{Buckley:2014ana} from the NNPDF3.0~\footnote{Approximate $\text{N}^3$LO PDF sets have become available recently. Given the time-scale of the present computation, and to compare to Ref.~\cite{ATLAS:2021mbt} we still use the NNPDF3.0 PDF set. As the main focus of this article is perturbative convergence and the technical challenges, we believe predictions based on recent PDFs will not change the message of this work.} PDF set~\cite{NNPDF:2014otw}, which was also used for the theory predictions in Ref.~\cite{ATLAS:2021mbt}. We set the factorization and renormalization scales to $\mu_F = \mu_R = m_{\gamma\gamma}$ and use standard seven-point scale variation to obtain uncertainty bands. As in Ref.~\cite{ATLAS:2021mbt}, we perform the calculation for the LHC at 13~TeV with fiducial cuts $p_{T,\gamma_{1(2)}}>40\,(30)$~GeV, $|\eta_\gamma|<2.37$ excluding $1.37 < |\eta_\gamma| < 1.52$, $\Delta R_{\gamma\gamma}>0.4$, using a hybrid scheme for photon isolation~\cite{Siegert:2016bre,Chen:2019zmr} which combines the smooth-cone condition
\begin{equation}
    E_T^\text{max}(r) = 0.15\, E_{T,\gamma}\, \frac{1-\cos(r)}{1-\cos(R^s_\text{max})},\ r<R^s_\text{max},
\end{equation}
with the hard-cone condition
\begin{equation}
    E_T^\text{max}(r) = 0.09\, E_{T,\gamma},\ r<R^h_\text{max},
\end{equation}
where $E_T^\text{max}(r)$ is the maximal allowed sum of transverse energies of all partons within angular distance $r$ of a photon, and we use $R^s_\text{max}=0.1$ and $R^h_\text{max}=0.2$.

\section{Six-point One-Loop Amplitudes}
As announced in the previous section, we obtain analytic expressions for the scattering amplitudes of the processes $0\to \gamma \gamma g g q \bar{q}$ and $0\to \gamma \gamma Q\bar{Q} q \bar{q}$ at one-loop order using finite-field-based reconstruction. While an analytic derivation of a six-point one-loop amplitude has already been performed using high-precision floating-point evaluations of the amplitude~\cite{Laurentis:2019bjh}, the present calculation constitutes the first full calculation of such an amplitude using finite-field reconstruction~\footnote{Note that in Ref.~\cite{Sarandrea:2023cfh}, the MHV 6-gluon amplitude has been computed in terms of momentum-twistor variables using finite-field reconstruction. We find, however, that the expressions of more complicated helicity configurations of our desired amplitude are too complicated when expressed through momentum-twistor variables, making a reconstruction unfeasible.}.~\footnote{This is also the first independent application of the methods from Ref.~\cite{DeLaurentis:2022otd}, in contrast to the previous computations involving at least one of its authors.}

The method of rational reconstruction over finite fields \cite{vonManteuffel:2014ixa, Peraro:2016wsq} consists of two steps. First, we develop a framework to calculate the amplitude numerically for arbitrary finite-field phase-space points, and second, we construct compact rational ans\"atze for the analytic expressions. The unknown coefficients in the latter can be obtained given a sufficient number of random finite-field probes.

Finite-field reconstruction is routinely used in the literature to calculate analytic expressions of two-loop five-point QCD amplitudes. In the five-point case, the ans\"atze can be built using Mandelstam variables and a single parity-odd quantity, which has been done to obtain massless~\cite{Abreu:2019odu,Badger:2019djh,Abreu:2020cwb,Chawdhry:2020for,Abreu:2021oya,Chawdhry:2021mkw} and massive~\cite{Abreu:2021asb,Badger:2021nhg,Hartanto:2022qhh} amplitudes. This is impractical for a generalization to six-point kinematics because the Gram determinant $G(p_1,p_2,p_3,p_4,p_5,p_6)$ is non-zero when expressed through Mandelstam variables but evaluates to zero in four dimensions. A solution is given by intrinsically four-dimensional momentum-twistor variables~\cite{Hodges:2009hk}, which have been used for the finite-field reconstruction of massless~\cite{Badger:2017jhb,Abreu:2017hqn,Badger:2018enw,Abreu:2018zmy,Badger:2021imn,Badger:2023mgf} and massive~\cite{Badger:2021ega,Badger:2022ncb,Badger:2024sqv,Badger:2024mir,Badger:2024dxo,Badger:2025ljy} five-point two-loop amplitudes as well. Momentum-twistor variables, however, have the disadvantage that they do not have a simple connection to physical limits and are related to Mandestam variables via rational---not polynomial---identities, introducing additional denominators. Indeed, Ref.~\cite{Abreu:2019odu} has found significantly simplified results when Mandelstam variables are used compared to twistor variables used in Ref.~\cite{Abreu:2018zmy}. Alternatively, phase-space variables which are both four-dimensional and have an intuitive behavior in infrared limits are given by spinor products in the spinor-helicity formalism. Their disadvantage is that standard finite-field reconstruction algorithms require a set of independent variables, whereas spinor products are related via Schouten and momentum-conservation identities. This can be tackled with methods from algebraic geometry as proposed in Ref.~\cite{DeLaurentis:2022otd}, which have been applied to the calculation of a number of massless~\cite{Abreu:2023bdp,DeLaurentis:2023nss,DeLaurentis:2023izi} and massive~\cite{Campbell:2022qpq,Campbell:2024tqg,DeLaurentis:2025dxw,Campbell:2025ftx,Campbell:2025llb,DeLaurentis:2026brm} five-point amplitudes. There, the finite-field probes were obtained either from previous analytic computations, through the unitarity method~\cite{Bern:1994zx,Ellis:2007br,Giele:2008ve}, or its implementation in the software package \texttt{Caravel}~\cite{Abreu:2020xvt}. However, the latter currently only supports the calculation of finite-field probes of amplitudes for five-point kinematics.

The reconstructed expressions of the amplitude have been made publicly available \cite{zenodo}. Details of the computation and subsequent implementation in efficient \texttt{C++} code can be found in the supplementary material \cite{suppl} to this letter (see also references \cite{Vermaseren:2000nd,Klappert:2019emp, Klappert:2020aqs,Chen:2019wyb,Maierhofer:2017gsa, Klappert:2020nbg,Henn:2022ydo,Chawdhry:2023yyx,DeLaurentis:2023qhd,flint,eschmentthesis} therein).

\section{Results}

\begin{figure*}
    \centering
    \includegraphics[width=\textwidth]{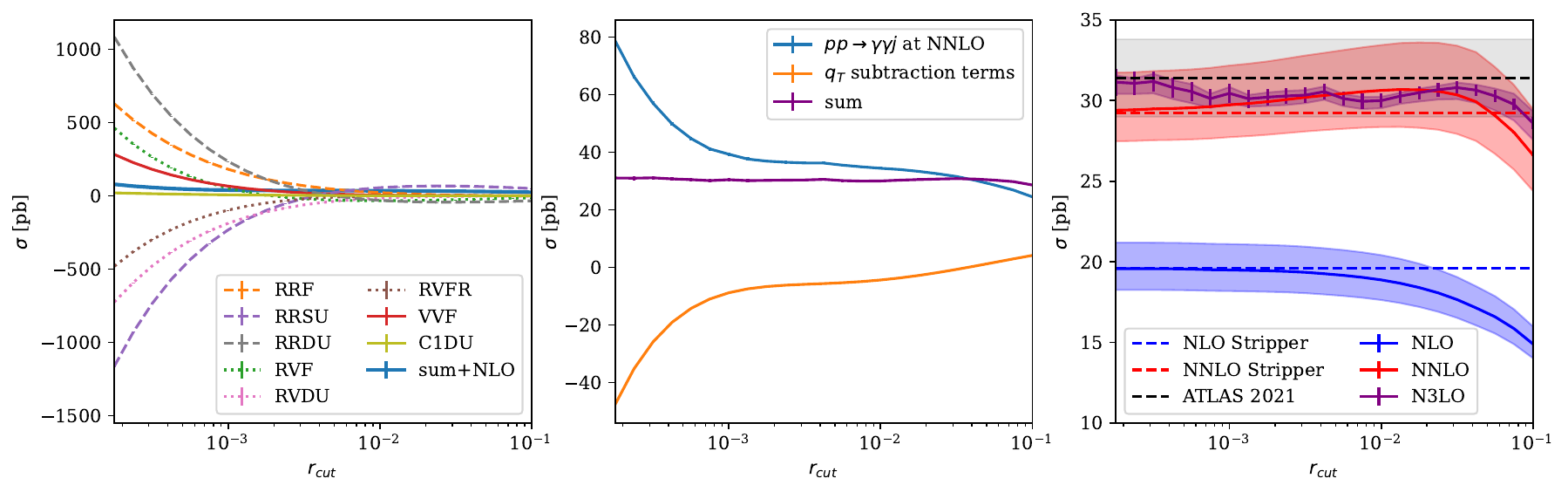}
    \caption{The left plot shows $pp\to \gamma\gamma j$ cross sections at NNLO for different $r_\text{cut}$ values decomposed into multiple contributions (a precise definition can be found in Ref.~\cite{Czakon:2019tmo}). The first two letters indicate contributions which are double-virtual (VV), real-virtual (RV), double-real (RR), or a single-convolution with splitting kernels (C1). The following letters correspond to finite (F), single-unresolved (SU), double-unresolved (DU), and finite-remainder (FR) contributions. The NNLO cross section and the contribution of the $q_T$ subtraction terms, Eq.~\eqref{eq:qt-slicing}, is shown in the middle plot. Their sum converges to $\sigma^{pp\to \gamma\gamma}_{\text{N}^3\text{LO}}$ as $r_\text{cut}\to 0$. The right plot displays the results of the $q_T$-slicing method additionally for NLO and NNLO, where the cross section can also be directly calculated with \texttt{STRIPPER} for comparison. It also contains a comparison to experimental data presented by ATLAS in Ref.~\cite{ATLAS:2021mbt}. The error bars correspond to the statistical uncertainty of the Monte Carlo integration, whereas the shaded bands indicate the uncertainty from seven-point scale variation for the theory predictions and the experimental uncertainty reported by ATLAS, respectively. The numerical values to reproduce the right plot can be found in the supplementary material to this letter.}
    \label{fig:xsec-rcut-dependence}
\end{figure*}
Our main result is the behavior of the fiducial $pp\to\gamma\gamma$ cross section at $\text{N}^3$LO for the slicing variable $r_\text{cut} \equiv q_T^\text{cut}/m_{\gamma\gamma} \ll 1$ shown in Fig.~\ref{fig:xsec-rcut-dependence}. The left plot demonstrates the drastic growth of various contributions to the NNLO $pp\to\gamma\gamma j$ cross section for small $r_\text{cut}$, which results in large cancellations between real and virtual corrections as well as integrated subtraction terms. Not illustrated are cancellations within each contribution~\cite {Janssen:2025zke}. These various cancellations are responsible for the substantial computational resources used---in units of million~CPUh: 2.7 for the real-virtual-finite (RVF), 2.3 for the double-real-finite (RRF), 2.4 for the double-real-single-unresolved (RRSU), 0.5 for the double-real-double-unresolved (RRDU), and 0.6 for the double-virtual-finite (VVF) contribution (other contributions required significantly less).

The middle plot of Fig.~\ref{fig:xsec-rcut-dependence} presents the cancellation between the two terms in Eq.~\eqref{eq:qt-slicing}. We observe that the $r_\text{cut}$ dependence of the sum is smaller than its statistical uncertainty over a large range. Furthermore, the separate divergence of the terms is clearly visible for $r_\text{cut}<10^{-3}$.

Finally, the right plot of Fig.~\ref{fig:xsec-rcut-dependence} shows the results for the fiducial $\text{N}^3$LO cross section, including the statistical error and the theoretical uncertainty from scale dependence, estimated with a seven-point variation. Although the scale dependence is affected by residual statistical uncertainty, it remains significantly smaller than the 8\% at NNLO. In view of the weak $r_\text{cut}$ dependence for small cutoff values, and the comparison of slicing and direct calculation at NNLO, any remaining dependence on the slicing parameter is expected to lie well within the statistical uncertainty. For our fiducial cross-section prediction, we choose $r_\text{cut}=3.2\times 10^{-4}$ for which the power corrections at NNLO amount to 0.3~pb.

As long as the N${}^3$LO power corrections are hidden in the statistical uncertainty band of $\pm 0.6$~pb, they too should not exceed $0.6$~pb. Estimating them at twice the power corrections of the NNLO result should be considered conservative. There have been attempts to remove at least the linear power corrections in $r_\text{cut}$ due to the asymmetric cuts on the two-body phase space \cite{Buonocore:2021tke}. Unfortunately, Ref.~\cite{Buonocore:2021tke} shows that precisely in the case of diphoton production, this leads to even larger differences between the exact NNLO cross section and the one obtained with $q_T$-slicing. In Ref.~\cite{Campbell:2024hjq}, it has been pointed out that the problem is related to the very specific photon-isolation criteria. The remedy proposed there seems to improve the situation slightly, but is based on a special treatment of only some of the contributions at NNLO and does not seem to be extensible to N${}^3$LO in a justifiable way. Hence, for now, the only way to deal with power corrections at this order is to take as small $r_\text{cut}$ as possible.

Our result reads 
\begin{equation}
    \sigma^{pp\to\gamma\gamma}_{\text{N}^3\text{LO}} = 31.2(6)^{+0.5}_{-0.7}~\text{pb} \, ( \pm 0.6 \text{ pb power corrections}) ,
\end{equation}
where the statistical uncertainty is quoted in parentheses, the scale uncertainty is provided as lower and upper bounds of the variation range, and the power-corrections estimate is separated. The scale uncertainty is obviously affected by the statistical uncertainty. Thus, we recommend adding the uncertainties in quadrature to obtain the total error estimate. In summary, our result reduces the uncertainty on the total cross section to 3\%, while also being consistent with the NNLO value, indicating convergence of the perturbation series. Additionally, it is in agreement with the ATLAS measurement~\cite{ATLAS:2021mbt}
\begin{equation}
    \sigma^{pp\to\gamma\gamma}_{\text{ATLAS}} = 31.4\pm0.1\text{ (stat.)}\pm2.4\text{ (syst.) pb}
\end{equation}
also displayed in Fig.~\ref{fig:xsec-rcut-dependence}.

\begin{figure}
    \centering
    \includegraphics[width=\linewidth]{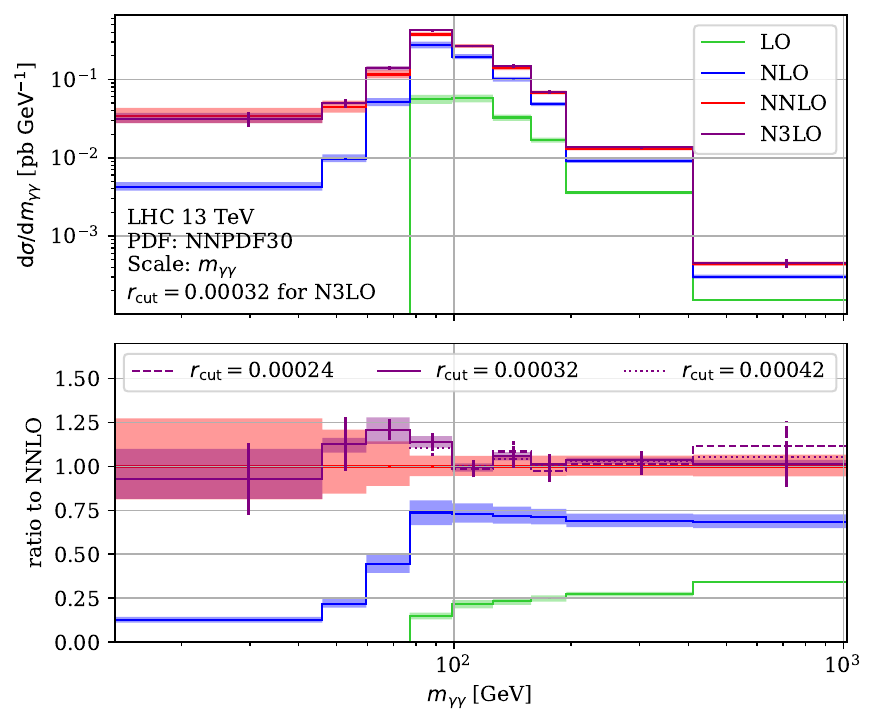}
    \caption{The invariant-mass distribution of the photon pair through $\text{N}^3$LO. The top panel contains absolute predictions, whereas the bottom panel is normalized to the NNLO values. While the distributions through NNLO are calculated through direct computation with \texttt{STRIPPER}, the $\text{N}^3$LO values are obtained through $q_T$ slicing for fixed $r_\text{cut}$. The remaining dependence on $r_\text{cut}$ is included in the lower panel and lies well within the statistical uncertainty.}
    \label{fig:m_yy_dist}
\end{figure}

Lastly, since $q_T$ slicing yields fully differential information, we can obtain distributions for arbitrary observables. Fig.~\ref{fig:m_yy_dist} shows the differential cross section $\textrm{d}\sigma^{pp\to\gamma\gamma}/\textrm{d}m_{\gamma\gamma}$ through $\text{N}^3\text{LO}$. As we have optimized the phase-space integration to minimize the statistical uncertainty of the total cross section, there remain large error bars for $m_{\gamma\gamma}$ regions that contribute less. Overall, the statistical uncertainty is comparable with the scale uncertainty at NNLO. Nevertheless, the distribution shows perturbative convergence across different regions of the phase space, as well as stable behavior as the value of the slicing parameter $r_\text{cut}$ is varied.

\section{Conclusions}

Theoretical predictions for $pp \to \gamma\gamma$ are plagued by large higher-order corrections, but our $\text{N}^3$LO result finally shows perturbative convergence.
The $\text{N}^3$LO corrections are small, in agreement with the NNLO QCD uncertainty estimates, and lead to a significant reduction of the scale sensitivity. They shift the theoretical prediction slightly upward, closer to the central value of the ATLAS measurement, although this is of little relevance given the large experimental systematic uncertainty. Since the calculation has been performed in a fully differential manner, extending to different phase spaces or observables based on the photons' kinematics is straightforward.

This is a first $\text{N}^3$LO QCD calculation for a process with actual $2\to2$ kinematics.
It marks a milestone in the advancement and automation of higher-order predictions, specifically at NNLO. Besides technical improvements of the cross-section integration, it required substantial effort to increase the stability and evaluation speed of the one-loop matrix elements.

Despite all improvements and considerable computational resources used, the precision of the calculated $\text{N}^3$LO QCD cross sections is limited by the statistical uncertainty arising from the Monte Carlo integration of the NNLO QCD cross section of the $pp \to \gamma\gamma j$ process in the limit of vanishing photon-pair transverse momentum. The cancellation between positive and negative contributions in subtracted cross sections is a major limiting factor.
Improving on this issue would make a substantial difference in the capabilities of existing NNLO QCD frameworks and their application to $\text{N}^3$LO slicing computations.
Turning this argument around, local $\text{N}^3$LO subtraction schemes might offer improved efficiency by avoiding regions with extreme numerical cancellations. Furthermore, they would remove the uncertainty due to power corrections.

Based on our results, one can claim that diphoton production at N${}^3$LO has a theoretical uncertainty of 2\% as long as the computational methodology itself does not introduce additional errors. This should be sufficient to improve various measurements by constraining Monte Carlo simulations currently used.

\section*{Acknowledgments}
This work was performed in part using the Cambridge Service for Data Driven Discovery (CSD3), part of which is operated by the University of Cambridge Research Computing on behalf of the STFC DiRAC HPC Facility (www.dirac.ac.uk). The DiRAC component of CSD3 was supported by STFC grants ST/P002307/1, ST/R002452/1 and ST/R00689X/1. The authors gratefully acknowledge the computing time provided to them at the NHR Center NHR4CES at RWTH Aachen University (project number \texttt{p0020025} and \texttt{p0020216}). This is funded by the Federal Ministry of Research, Technology and Space, and the state governments participating on the basis of the resolutions of the GWK for national high performance computing at universities. The work of M.C.\ and F.E.\ was supported by the Deutsche Forschungsgemeinschaft (DFG) under grant 396021762 - TRR 257: Particle Physics Phenomenology after the Higgs Discovery, and grant 400140256 - GRK 2497: The Physics of the Heaviest Particles at the LHC. T.G. has been supported by STFC consolidated HEP theory grants ST/T000694/1 and ST/X000664/1. R.P.\ acknowledges that this research was funded in part by NCN 2024/55/D/ST2/00934.

\bibliography{apssamp}% Produces the bibliography via BibTeX.

\end{document}

\title{Supplementary Material}% Force line breaks with \\

\date{\today}% It is always \today, today,
             %  but any date may be explicitly specified

%\keywords{Suggested keywords}%Use showkeys class option if keyword
                              %display desired
\maketitle

\section{Amplitude Calculation}

In this section, we give supplementary information on the calculation of the one-loop six-point amplitudes of the processes $0\to \gamma \gamma g g q \bar{q}$ and $\quad 0\to \gamma \gamma Q\bar{Q} q \bar{q}$. As explained in the main text, the calculation consists of two steps: obtaining numerical finite-field probes and reconstructing a compact analytic solution.

For the first step, we generate Feynman diagrams with the private software \texttt{DiaGen}, manipulate them in \texttt{FORM}~\cite{Vermaseren:2000nd} to perform color and Dirac algebra, and export them as \texttt{C++} functions. Finite-field variables are represented in the \texttt{FFInt} data type from \texttt{FireFly}~\cite{Klappert:2019emp, Klappert:2020aqs}. Within \texttt{C++}, we apply projectors to helicity amplitudes following Ref.~\cite{Chen:2019wyb} and express scalar products involving the loop momentum as inverse propagators. The resulting loop integrals are reduced to master integrals by linking with \texttt{Kira}~\cite{Maierhofer:2017gsa, Klappert:2020nbg}, which internally performs integration-by-part reduction in the \texttt{FFInt} data type. Finally, we map the master integrals to the function basis used in  Ref.~\cite{Henn:2022ydo} employing the solution provided in their ancillary file. In summary, we develop a framework to compute finite-field values of the rational coefficients of all basis functions appearing in the helicity amplitudes, given a phase-space point in the finite field.

For the second step, we closely follow the procedure outlined in Ref.~\cite{DeLaurentis:2022otd}. We use ans\"atze consisting of the spinor products
\begin{equation}
    \langle ij\rangle \text{ and } [ij],\quad \text{where } 1\leq i< 6,\ i<j\leq 6.
\end{equation}
We probe the rational coefficients in the amplitude given finite-field values for these spinor products~\footnote{Only 14 of the 30 variables can be chosen independently. The others are constrained by momentum conservation and Schouten identities.}. Furthermore, we can use probes on $p$-adic fields by converting $p$-adic numbers to integers, reconstructing the result as a rational number using \texttt{FireFly}, and converting the rational number back to a $p$-adic number (see Ref.~\cite{Chawdhry:2023yyx} for details). In accordance with the letters of the function alphabet of the integrals \cite{Henn:2022ydo}, the denominators of the rational coefficients can only contain factors contained in the set
\begin{equation}
    D\equiv\{\langle ij\rangle,[ij],s_{ijk},\langle i|j+k|i],\langle i|j+k|l],\Delta_{ij|kl|mn}\},
\end{equation}
where $\langle i|j+k|l]\equiv \langle ij\rangle[jl]+\langle ik\rangle [kl]$ and $\Delta_{ij|kl|mn}\equiv \frac{1}{4}\lambda(s_{ij},s_{kl},s_{mn})$ with the K\"all\'en function $\lambda(x,y,z)\equiv x^2+y^2+z^2-2(xy+yz+zx)$, and the indices $i,\ldots, n$ run from 1 through 6. This enables us to fully determine the denominators from $p$-adic probes on the varieties $V(\langle D_k \rangle_{R_6})$, associated to the codimension-one ideals $\langle D_k \rangle_{R_6}$ generated by the potential denominator factors $D_k$~\footnote{For the relevant definitions, we refer to Ref.~\cite{DeLaurentis:2022otd}, where we also borrow the notation.}. We generate two-digit $p$-adic values for four different $p\sim 10^3$ for the spinor products using the software package \texttt{lips}~\cite{DeLaurentis:2023qhd} and extract the denominator power from the $p$-adic valuation computed with \texttt{FLINT}~\cite{flint}, demanding that at least three of the four valuations agree~\footnote{Due to the relatively small size of $p$, it can happen by chance that a non-zero number is mapped to zero, impacting the valuation.}. After determining the denominator (and potentially some numerator factors), one can obtain a finite-sized ansatz for the amplitude due to the fixed mass dimension and little-group weight. Given a sufficient number of finite-field probes, an analytic expression for the amplitude can then, in principle, be obtained by reconstructing the rational numbers in the ansatz. To make this reconstruction feasible and to obtain a more compact final result, we simplify the ansatz in multiple steps:
\begin{enumerate}
    \item Numerically, find linear relations between rational coefficients and use them to eliminate coefficients where the numerator has a large mass dimension.
    \item Study relations between coefficients in the limits $\langle i|j+k|l]\to 0$ or $\Delta_{ij|kl|mn}\to 0$. Since they do not directly correspond to infrared limits, poles in these variables are spurious and have to cancel between terms in the amplitude. This can be used to obtain leading-pole contributions to complicated coefficients from terms that are easier to obtain.
    \item Apply partial fractioning to the remaining coefficients. Valid partial-fraction decompositions can be obtained from $p$-adic probes on varieties associated to codimension-two prime ideals. We use six-point primary decompositions provided in Appendix~B of Ref.~\cite{DeLaurentis:2025dxw}.
\end{enumerate}
For a more detailed description of the reconstruction procedure used in this calculation, see Ref.~\cite{eschmentthesis}.

\begin{figure}
    \centering
    \includegraphics[width=.4\linewidth]{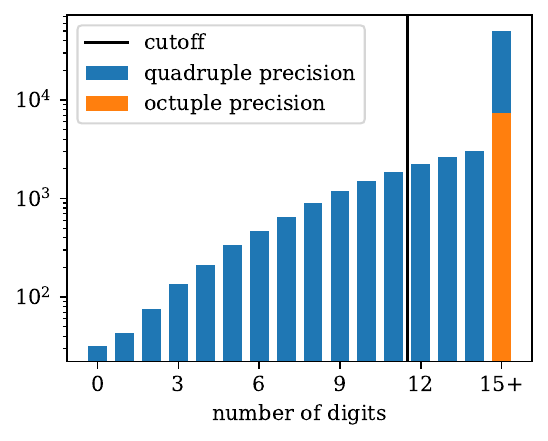}
    \caption{Distribution of the number of corrects digits for a sample of evaluations of six-point one-loop amplitudes. The cutoff indicates that we demand at least 12 correct digits before switching to octuple precision, which always yields at least 15 digits.}
    \label{fig:amplitudes_precision}
\end{figure}
We implement the obtained amplitude in \texttt{C++} code using \texttt{FORM} output optimization and compile it through octuple precision with the \texttt{qd} library~\cite{qd-library}. For every evaluation, we reevaluate the amplitude after rescaling all momenta and the renormalization scale by a factor of 10. From the difference between the expected and obtained rescaled amplitude values, we can estimate the number of correct digits. Since we constantly evaluate the amplitude in infrared phase-space regions, we find that double precision is usually insufficient. Therefore, we use quadruple precision by default. As shown in Fig.~\ref{fig:amplitudes_precision}, this yields more than 15 correct digits in most cases. However, some evaluations drop in precision, and if we estimate fewer than 12 correct digits, we reevaluate the amplitude in octuple precision, which is always sufficient. The average evaluation time is 0.5~seconds per phase space point.
	
\section{Improved selector function}
The sector-improved residue subtraction scheme employed in this work to handle infrared singularities in real-emission contributions relies on a sector decomposition of the phase space. This sector decomposition guarantees that only a small subset of all infrared singularities needs to be considered at a time, and is achieved using a set of selector functions. A selector function should approach zero in all singular limits, except for the small subset of singular limits it is supposed to select. The sector decomposition is then achieved by multiplying the fully differential cross section by the different selector functions. Furthermore, summing over all selector functions should yield unity. This way, summing over all sectors recovers the full cross section. At NNLO, two types of selector function are needed, corresponding to the two types of sectors: triple-collinear sectors, which contain configurations where three partons are simultaneously collinear, and double-collinear sectors, which contain configurations where two distinct pairs of partons become collinear at the same time. Previously, the set of selector functions proposed in Ref.~\cite{Czakon:2014oma} were used, which read
\begin{equation}
    \mathcal{S}_{ij,k} = \frac{1}{D_2\:d_{ij,k}}\;,\;\;\;\;\mathcal{S}_{i,k;j,l} = \frac{1}{D_2\:d_{i,k}\:d_{j,l}}\;,\;\;\;\;D_2 = \sum_{ij}\bigg[\sum_k\frac{1}{d_{ij,k}}+\sum_{kl}\frac{1}{d_{i,k}\:d_{j,l}}\bigg]\;,
\end{equation}
\begin{equation}
d_{i,k} = \bigg(\frac{E_i}{\sqrt{s}}\bigg)^\alpha (1-\cos \theta_{ik})^\beta\;,\;\;\;\;d_{ij,k} = \bigg(\frac{E_i}{\sqrt{s}}\frac{E_j}{\sqrt{s}}\bigg)^\alpha \big[(1-\cos \theta_{ij})(1-\cos \theta_{ik})(1-\cos \theta_{jk})\big]^\beta\;,
\end{equation}
where $\mathcal{S}_{ij,k}$ selects the triple-collinear sector involving partons $i$, $j$, and $k$ where $i$ and $j$ are allowed to become soft, and $\mathcal{S}_{i,k;j,l}$ selects the double-collinear sector involving collinear singularities between partons $i$ and $k$ and partons $j$ and $l$ and soft singularities involving partons $i$ and $j$. $\alpha$ and $\beta$ are free parameters, but must be chosen to be larger than zero.

This form of the selector functions is sufficient, but numerically suboptimal in certain pathological kinematic configurations. The problem is related to configurations where four partons are simultaneously close to collinear. While four partons cannot be exactly collinear at NNLO, the calculation presented in this letter for the production of a photon pair at non-zero transverse momentum can reach very small values of $r_\textrm{cut}$, thus allowing for a parton to be nearly collinear to the initial state at leading order. This implies that the NNLO calculation will contain configurations with four nearly collinear partons. In such configurations, the form of the selector functions chosen in Ref.~\cite{Czakon:2014oma} disproportionately favours triple-collinear sectors, leading to an unnecessarily strongly peaked integrand. To see this, consider the situation where all four nearly collinear partons are separated from each other by angles of the order of $\delta\ll1$. In this case, a double-collinear selector function $\mathcal{S}_{i,k;j,l}$ involving all four partons is suppressed relative to a triple-collinear selector function involving three of them by $\mathcal{O}(\delta^{2\beta})$. If, starting from this configuration, the double-collinear singularity is then approached by reducing $\theta_{ik}$ and $\theta_{jl}$ while keeping the other angles $\mathcal{O}(\delta)$, the triple-collinear sector only starts being shielded from this singularity once $\theta_{ik}$ and $\theta_{jl}$ are $\mathcal{O}(\delta^2)$. In short, the old set of selector functions sometimes allow a double-collinear singularity to be approached very closely in triple-collinear sectors, leading to an extremely strongly peaked integrand, dramatically slowing down the Monte Carlo convergence. Furthermore, this same behaviour also increases the size of individual cross section contributions, leading to larger cancellations in the final result, further slowing down the numerical convergence. A similar problem occurs if three partons $(i,j,k)$ are close to collinear at the same time two other partons $(l,m)$ with $l,m\notin\{i,j,k\}$ are also nearly collinear.

To resolve this problem, a new selector function is introduced for triple-collinear sectors which involves only two angles, rather than three. Explicitly, we use
\begin{equation}\label{eq:newSelectorFunction}
d_{ij,k} = \bigg(\frac{E_i}{\sqrt{s}}\frac{E_j}{\sqrt{s}}\bigg)^\alpha \big[(1-\cos \theta_\textrm{min})(1-\cos \theta_\textrm{max})\big]^\beta\;,\;\;\;\;\theta_\textrm{min/max} = \min/\max \{ \theta_{ij},\theta_{ik},\theta_{jk}\}\;.
\end{equation}
In the configuration discussed above, triple-collinear and double-collinear selector functions are now on equal footing: neither is suppressed relative to the other by factors of $\delta$. As a result, unnecessarily strongly peaked integrands are avoided and the cancellation between different contributions is reduced. We estimate that this simple change speeds up the Monte Carlo convergence by roughly a factor of 3000.

Note that the definition in Eq.~\eqref{eq:newSelectorFunction} is not the only possibility. As always, there is a lot of freedom when fixing the selector function. In particular, it does not need to be proportional to a product of two factors like $(1-\cos \theta)^\beta$. Even when limiting oneself to such forms, while the presence of $\theta_\textrm{min}$ is necessary to properly capture single-collinear configurations, $\theta_\textrm{max}$ can be replaced with e.g.~$\theta_\textrm{mid}$, i.e.~the angle that is neither the smallest nor the largest, or $\theta_\textrm{mean}=(\theta_{ij}+\theta_{ik}+\theta_{jk})/3$, etc.

\section{Numerical Cross Section Results}
In the right plot of Fig.~2 of the paper, the $pp\to \gamma\gamma$ cross section at the LHC is displayed at different orders in QCD. Here, we report the numerical values with which one can reproduce the plot. Using direct calculation in \texttt{STRIPPER}, one obtains $\sigma_\text{NLO}=19.62^{+1.62}_{-1.32}$ and $\sigma_\text{NNLO}=29.21(5)^{+2.28}_{-1.89}$. The values obtained with $q_T$ slicing are displayed in Table~\ref{tab:r-dependence}.

\begin{table}
    \centering
    \caption{Values obtained with $q_T$ slicing that are displayed in the right plot of Fig.~2 of the paper.}
    \begin{tabular}{c@{\hskip 1cm}c@{\hskip 1cm}c@{\hskip 1cm}c}
    \hline
    $r_\text{cut}$ & $\sigma_\text{NLO}$ & $\sigma_\text{NNLO}$ & $\sigma_{\text{N}^3\text{LO}}$ \\\hline
        0.000178 & $19.57(3)^{+1.62}_{-1.32}$ & $29.40(9)^{+2.33}_{-1.93}$ & $31.14(83)^{+0.43}_{-0.72}$ \\[.2cm]
0.000237 & $19.58(3)^{+1.62}_{-1.32}$ & $29.43(9)^{+2.33}_{-1.93}$ & $31.07(70)^{+0.31}_{-0.66}$ \\[.2cm]
0.000316 & $19.57(3)^{+1.62}_{-1.32}$ & $29.48(8)^{+2.35}_{-1.94}$ & $31.19(64)^{+0.48}_{-0.74}$ \\[.2cm]
0.000422 & $19.56(2)^{+1.62}_{-1.32}$ & $29.51(7)^{+2.36}_{-1.95}$ & $30.81(61)^{+0.46}_{-0.72}$ \\[.2cm]
0.000562 & $19.54(2)^{+1.61}_{-1.31}$ & $29.56(7)^{+2.38}_{-1.96}$ & $30.56(58)^{+0.46}_{-0.73}$ \\[.2cm]
0.000750 & $19.51(2)^{+1.61}_{-1.31}$ & $29.63(7)^{+2.41}_{-1.98}$ & $30.12(60)^{+0.30}_{-0.57}$ \\[.2cm]
0.001000 & $19.49(2)^{+1.61}_{-1.31}$ & $29.74(7)^{+2.44}_{-2.01}$ & $30.45(59)^{+0.40}_{-0.62}$ \\[.2cm]
0.001334 & $19.47(2)^{+1.60}_{-1.31}$ & $29.82(7)^{+2.47}_{-2.03}$ & $30.11(57)^{+0.24}_{-0.49}$ \\[.2cm]
0.001778 & $19.44(2)^{+1.60}_{-1.30}$ & $29.93(7)^{+2.50}_{-2.05}$ & $30.21(50)^{+0.29}_{-0.54}$ \\[.2cm]
0.002371 & $19.41(2)^{+1.60}_{-1.30}$ & $30.06(7)^{+2.54}_{-2.08}$ & $30.29(47)^{+0.32}_{-0.56}$ \\[.2cm]
0.003162 & $19.35(2)^{+1.59}_{-1.30}$ & $30.20(6)^{+2.59}_{-2.12}$ & $30.34(46)^{+0.31}_{-0.50}$ \\[.2cm]
0.004217 & $19.26(2)^{+1.58}_{-1.29}$ & $30.32(6)^{+2.64}_{-2.15}$ & $30.55(46)^{+0.33}_{-0.50}$ \\[.2cm]
0.005623 & $19.16(2)^{+1.57}_{-1.28}$ & $30.44(6)^{+2.70}_{-2.19}$ & $30.13(45)^{+0.28}_{-0.47}$ \\[.2cm]
0.007499 & $19.04(2)^{+1.56}_{-1.26}$ & $30.55(6)^{+2.75}_{-2.23}$ & $29.95(49)^{+0.27}_{-0.48}$ \\[.2cm]
0.010000 & $18.88(2)^{+1.54}_{-1.25}$ & $30.63(6)^{+2.81}_{-2.27}$ & $30.00(43)^{+0.24}_{-0.47}$ \\[.2cm]
0.013335 & $18.67(2)^{+1.51}_{-1.23}$ & $30.69(6)^{+2.88}_{-2.32}$ & $30.28(43)^{+0.21}_{-0.48}$ \\[.2cm]
0.017783 & $18.40(2)^{+1.48}_{-1.20}$ & $30.66(6)^{+2.94}_{-2.36}$ & $30.49(41)^{+0.25}_{-0.55}$ \\[.2cm]
0.023714 & $18.07(1)^{+1.45}_{-1.17}$ & $30.58(6)^{+3.00}_{-2.39}$ & $30.67(41)^{+0.25}_{-0.61}$ \\[.2cm]
0.031623 & $17.65(1)^{+1.40}_{-1.14}$ & $30.35(6)^{+3.05}_{-2.43}$ & $30.80(37)^{+0.32}_{-0.71}$ \\[.2cm]
0.042170 & $17.14(1)^{+1.34}_{-1.09}$ & $29.93(5)^{+3.08}_{-2.43}$ & $30.65(36)^{+0.34}_{-0.80}$ \\[.2cm]
0.056234 & $16.59(1)^{+1.28}_{-1.04}$ & $29.07(5)^{+3.01}_{-2.38}$ & $30.28(37)^{+0.47}_{-0.89}$ \\[.2cm]
0.074989 & $15.86(1)^{+1.20}_{-0.97}$ & $28.03(5)^{+2.94}_{-2.31}$ & $29.76(34)^{+0.67}_{-1.00}$ \\[.2cm]
0.100000 & $14.90(1)^{+1.09}_{-0.88}$ & $26.63(5)^{+2.85}_{-2.23}$ & $28.61(36)^{+0.74}_{-1.04}$ \\\hline
    \end{tabular}
    \label{tab:r-dependence}
\end{table}

\bibliography{apssamp}